\documentstyle[aps,epsf]{revtex}
\def\ul#1{\underline{#1}}
\def\kk{\ul{k}}

\def\mat#1{\ul{\ul{#1}}}
 \def\w{\omega}

\def\e{\varepsilon}

\def\dps{\displaystyle}

\def\bg{\begin{equation}}

\def\en{\end{equation}}

\def\e{\epsilon}

\def\w{\omega}

\def\bge{\begin{eqnarray}}
\def\ene{\end{eqnarray}}
\def\ul#1{\underline{#1}}

\def\kk#1{\ul{k_{#1}}}
\def\ul#1{\underline{#1}}
\def\kk{\ul{k}}
\def\kks{{\kk\sigma}}

\def\mat#1{\ul{\ul{#1}}}

\def\w{\omega}
\def\e{\varepsilon}

\def\ga{{\Gamma\alpha}}
\def\gapara#1#2{{\Gamma_{#1}\alpha_{#2}}}
\def\gpap{{\Gamma'\alpha'}}
\def\nn{\frac{1}{N_s}}

\begin{document}

 \twocolumn[\hsize\textwidth\columnwidth\hsize\csname
 @twocolumnfalse\endcsname

\title{Self-consistent Treatment of Crystal-Electric-Field-Levels  in the
Anderson Lattice}
	
\author{F.~B.~Anders and D.L.Cox}
\address{Department of Physics, The Ohio State University, 
 Columbus, Ohio, 43210-1106}
\address{Institute for Theoretical Physics, UCSB, Santa Barbara,
CA, 93106-4030}
\date{26.~June 1996}

\maketitle

\begin{abstract}
We consider an  Anderson lattice model with a  spin 1/2 degenerated
conduction electron band and localized ionic CEF-levels, classified
according to the irreducible representation of the point group of the
lattice. We present the self-consistency equations for 
local approximations ("$d\rightarrow\infty"$ approximation) for the
periodic Anderson model. It leads to
a matrix formulation of  the effective local density of states  and
the lattice $f$-Green's function. We derive the quasi-particle life-time
which  enters the Boltzmann transport equations. The impact 
of a $k$-dependent hybridization is discussed. 
We prove that vertex
corrections will vanish, as long as all states of an irreducible
representation couple to the conduction electron band with a
hybridization matrix element of the same parity. 
\end{abstract}

\vfill
\\
\vskip 3mm

]  

\narrowtext

\section{Introduction}
In Heavy Fermion (HF) materials \cite{Grewe91}, especially in Uranium
based compounds, a simple Anderson-lattice model with single $N$=fold
degenerate ionic ground-state, cannot explain the rich variety of
transport measurements in the whole  accessibly
temperature range. The  additional maxima found experimentally in the
specific heat, the resistivity and the thermo-power are related to
higher Crystal-Electric-Field (CEF) levels and have been used to
propose  CEF level schemes for many different HF compounds using
high temperature impurity approaches. 
Many theoretical attempts have been put forward to include CEF effects
in a many body description of HF materials \cite{Maekawa86}. For
magnetic impurities the 
high temperature spin-disorder resistance calculations
\cite{CornutCoqblin72} have been extended to describe anisotropy of
transport of CePt$_2$Si$_2$ \cite{Bhatta89} in third order perturbation
theory  neglecting  lattice coherence effects, 
for the low temperature phase it can included
within some limits in a non-crossing approximation (NCA) calculation
\cite{Bickers87,Strong94}.

 Nevertheless a theoretical approach for the lattice problem 
is missing. Additionally,
the low temperature scale $T^*$ will reflect higher
excited multiplets due to the remaining virtual fluctuations 
in the ground state. Since in Uranium based compounds the bare spectroscopic
CEF structure is still unsettled and a CEF singlet cannot be ruled out
as ground state a model for competition between CEF and Kondo singlet
was put forward only recently \cite{Kuramoto93}.
But for the Anderson lattice, approaches for including CEF effects
exist only within the slave boson mean field theory \cite{Evens92}.
The aim of this paper is to
present a formalism to incorporate CEF-levels in a dynamical mean
field (local) approximation \cite{Georges96}  which
is independent of the method used to solve the effective site
problem.  Even though we will restrict ourselves to one unoccupied state,
and  singly occupied CEF-multiplets
transforming according to different irreducible representations
$\Gamma_i$ of the point group of the lattice, we can extend our
method to address the question how to generate a multi-channel Kondo
lattice model from first principles.

\section{Theory}

It is assumed that a spin
degenerate conduction electron band couples to localized ionic states
via hybridization matrix elements. The matrix elements can be derived
by expanding the conduction band locally in the irreducible
representations of the point group \cite{CoqblinSchrieffer69}. First we
introduce the Hamiltonian and the notation used throughout the paper.
\begin{equation}
H =
\begin{array}[t]{l}
\dps
\sum_{\kks} \e_{\kks} c^\dagger_\kks c_\kks
+
\sum_{\nu\ga}E_\ga X_{\ga,\ga}^{\nu}
\\
\dps
+
\sum_{\nu,\ga,\kks}
V_{\ga}(\kks)e^{i\kk\ul{R}_\nu} c^\dagger_{\kks} X_{0,\ga}^\nu
+
h.c.

\end{array}
\label{eqn-1}
\end{equation}
The ionic states are labeled by their corresponding irreducible
representation of the point group and an index $\alpha$ which is a
shorthand notation for all other quantum numbers, like occupation number $n$,
spin, or orbit quantum number,
and $ X_{\gpap,\ga}^\nu = |\gpap><\ga|$ at the site
$\nu$. It is the most general formulation of the periodic 
Anderson model which takes into account only local hybridization. In
this paper we concentrate on local fluctuations between
singly and empty states only ($U\rightarrow \infty$). 


While propagating through the lattice, the conduction electrons are
scattered by the CEF states via the local $T$-matrix $V_{\ga}(\kks)
F_{\ga,\gpap}(z)V_{\gpap}(\kks)$, where
$F_{\ga,\gpap}(z) \equiv \ll X_{0,\ga}(\tau) X_{\gpap,0}\gg $ is the
local Green's function. In the summation of all scattering processes of the
conduction electrons, terms of the structure 
$$
\dps
\sum_{\ga} \sum_\sigma V^\star_{\gapara{1}{1}}(\kks)
\underbrace{\frac{1}{z-\e_{\kks}}}_{Band}  
\underbrace{V_{\ga}(\kks)
F_{\ga,\gapara{2}{2}}(z)V_{\gapara{2}{2}}(\kks)}_{T-matrix} 
$$ 
occur which can be identified as a component of a matrix product 
$
\left[
\mat{d}(\kk,z) \cdot \mat{F}(z)
\right]_{\gapara{1}{1},\gapara{2}{2}}
$
where we have defined the conduction electron matrix 
$$
 \mat{d}(\kk,z) =  \sum_\sigma  \mat{d}(\kk,\sigma,z) = \sum_\sigma
\ul{V}(\kks) \frac{1}{z-\e_{\kks}} \ul{V}^T(\kks)
$$ 
and 
$
\ul{V}^T(\kks) =
\left(
V^\star_{\gapara{1}{1}}(\kks),V^\star_{\gapara{1}{2}}(\kks), \cdots,
V^\star_{\gapara{n}{\Gamma_n}}(\kks) 
\right)
$. The f-Green's function $\mat{F}(z)$ is a $N\times N$ matrix, where $N$ is the
number of singly occupied states included in (\ref{eqn-1}).

{\bf The local ("$d\rightarrow\infty$") approximation:}\\
The local approximation \cite{Kuramoto87}, which is
equivalent to the limit $d\rightarrow\infty$ with an appropriate
rescaling of the effective hopping \cite{Georges96},
choose one $f$-site as an effective site which is embedded in an
effective medium Green's function generated self-consistently by the rest of
lattice. While in a single impurity problem the bare medium GF
$\mat{\Delta}_0(z) = \frac{1}{N_s}\sum_{\kk}\mat{d}(\kk,z)$
enters, the condition that the local $f$-Green's function has to be
equal to the $\kk$-summed lattice Green's function
\begin{equation}
\label{eqn-scc-doo}
 \nn \sum_{\kk}
\frac{1}{\mat{1}- \mat{\tilde F}(z)\left(\mat{d}(k,z) -
\mat{\tilde\Delta}(z)\right)}
= 1 \;\; .
\end{equation}
determined self-consistently the renormalized media
$\mat{\tilde\Delta}(z)$. The effective Anderson width
$\mat{\tilde\Gamma}(z) = \frac{\Im m}{\pi}\mat{\tilde\Delta}(z)$ and the
Green's function of the effective site $\mat{\tilde F}(z)$ are
block-diagonal in the irreducible representations of the point group,
since they are  local quantities.

{\bf The lattice Green's-functions}\\
The $f$-Green's-function (GF) matrix $\mat{F}(k,z)$ is obtain by summing over
all possible intermediate scattering events taking an electron from site
$i$ to $j$ and Fourier transforming  the result in the reciprocal lattice
space is
\begin{equation}
\mat{F}(k,z) 
=
\frac{1}{\mat{\tilde F}^{-1}(z) - \left(\mat{d}(\kk,z)
 -\mat{\tilde\Delta}(z) \right)} \;\;. 
\label{eqn-lnca-f}
\end{equation}
While the self-consistency condition (SCC) of the so-called lattice-NCA,
another well established 
local approximation for the Anderson lattice, is derived from a
different philosophy \cite{Grewe87}, the structure of the lattice
Green's function can be obtained from (\ref{eqn-lnca-f}) by replacing 
$\mat{\tilde\Delta}(z)$ by the bare $\mat{\Delta}_0(z)$.
Even though $\mat{\tilde F}(z)$ and $\mat{\tilde\Delta}(z)$ transform according
to the irreducible representations of the point group, the term
$\mat{d}(k,z)$ mixes different 
representations  for an arbitrary $k$-point destroying the point-group
symmetry in $k$-space.

Via the {\em exact} equation of motion
\begin{equation}
\begin{array}{rcl}
G_{\kks}(z) &=&\dps G_{\kks}^{(0)}(z) +  G_{\kks}^{(0)}(z) T_{\kks}(z)
G_{\kks}^{(0)}(z) 
\\
&\equiv &\dps
\frac{1}{ [G_{\kks}^{(0)}]^{-1}(z) - \Sigma_{\kks}(z)}
\end{array}
\end{equation}
and
\begin{equation}
 T_{\kks}(z) = 
\begin{array}[t]{l}
\dps
\sum_{\ga,\gpap} V_{\ga}(\kks) F_{\gpap,\ga}(\kk,z)V^\star_{\gpap}(\kks)
\\
= \ul{V}^{T}(\kks) \mat{F}(\kk,z) \ul{V}(\kks)
\end{array}
\end{equation}
a compact equation for the $k$-dependent self-energy
\begin{equation}
\Sigma_{\kks}(z) 
\begin{array}[t]{l}
=
\dps
\frac{ \ul{V}^{T}(\kks) \mat{F}(\kk,z) \ul{V}(\kks)}
{1 +  G_{\kks}^{(0)}(z) \ul{V}^{T}(\kks) \mat{F}(\kk,z) \ul{V}(\kks)}
\\[10pt]
\dps
= \ul{V}^{T}(\kks) \frac{1}{\mat{\tilde F}^{-1}(z) +
\mat{\tilde\Delta}(z) - \mat{d}(\kk,-\sigma,z)}
\ul{V}(\kks)
\end{array}
\label{eqn-sigma-c}
\end{equation}
is obtained from with the $k$-dependent inverse relaxation time
$\tau^{-1}(\kks,\w) = 2\Im m \Sigma_{\kks}(\w-i\delta)$ is calculate
entering Boltzmann transport theory. Even though a local approximation
has been used in derivating Eqn.(\ref{eqn-sigma-c}) the self-energy is
anisotropic due  the $k$-dependence of the hybridization.

{\bf Example: Two Kramers Doubles:}\\
The $4\times 4$ problem separates in two identical $2\times 2$
matrices for each pseudo-spin. Using the Faddeeva-function $w(z)$, the
diagonal elements $\tilde F_1(z)$ and $\tilde F_2(z)$ of the Green's
function matrix $\mat{F}(z)$ and angular averaged hybridization, we
obtain two SCC 
\begin{equation}
\begin{array}{l}
\label{eqn-ceff-d1}
\dps
\frac{1}{1 + \tilde F_\alpha(z)\tilde\Delta_\alpha(z)}
\left[1 - i\, w\left(\sqrt{\pi}\rho_0(z-\Sigma_{\sigma}(z)
-\e_0)\right) 
 \right . 
\\
\dps
\hspace{20mm}
\left.
\cdot
\frac{\pi V_\alpha^2\rho_0 \tilde F_\alpha(z)}{1 + \tilde
F_\alpha(z)\tilde \Delta_\alpha(z)}
\right]  =  1 
\hspace{5mm}
\alpha=1,2
\end{array}
\end{equation}
 which are only coupled via the self-energy of the conduction electrons
\begin{equation}
\Sigma_{\sigma}(z)= \sum_\alpha \frac{V_\alpha^2 \tilde
F_\alpha(z)}{1 + \tilde F_\alpha(z)\tilde\Delta_\alpha(z)} 
\;\; .
\label{equ-c-self}
\end{equation}
found from Eqn.(\ref{eqn-sigma-c}). 
The averaging of anisotropy effects is justified in the SCC, since only the 
effective density of states enters the local approximation. For the
transport calculation, however, the full angular dependence of the
hybridization gives rise to the anisotropy of the transport properties
even though a local approximation has been used. From
Eqn.~(\ref{equ-c-self}) it is clearly seen that the different position
of the Abricosov-Suhl resonaces in the two $\tilde F_\alpha(z)$ will
produced two maxima in the resistivity, as demonstrated in  \cite{Huth95}.

{\bf Vertex corrections in the Transport Theory:}\\
Normally a vertex function enters the calculation of transport
properties:
\begin{equation}
\begin{array}{rcl}
\dps
\Gamma_{\kk}(z, z+\nu)& =&\dps
 \partial_{\kk}\e_{\kk}
+ \sum_{\ul{k'}}  \partial_{\ul{k'}}\e_{\ul{k'}}G_{\ul{k'}}(z)
\\
&&
\cdot G_{\ul{k'}}(z+\nu) W_{\ul{k'},\ul{k}}(z,z+\nu) \;\;
\end{array}
\end{equation}
$W_{\ul{k'},\ul{k}}(z,z+\nu)$ being the irreducible two particle
propagator. If we restrict ourself to CEF-levels arising from the same
Hund's rule multiplet, the hybridization matrix element coupling both
conduction electron Green's function $G_{\ul{k'}}(z)$ to a local $f$
site will have 
the same parity. The total parity of the $\ul{k'}$-momentum loop is
given by $\partial_{\kk}\e_{\kk}$. Therefore the vertex correction will
be exactly zero in a lattice with inversion symmetry and
$\Gamma_{\kk}(z, z+\nu) = \partial_{\kk}\e_{\kk}$. 
Neverlethess, the total current has contributions from conduction
and $f$-electrons when a $\kk$-dependent hybridization is present
\cite{Leder81}. 

\section{Conclusion}
We have developed a formalism to include local CEF-levels in a
 local approximation. The lattice
coherence effects, which have been neglected in previous approaches
have been taken into account. Eqn.(\ref{eqn-lnca-f}) and
Eqn.(\ref{eqn-sigma-c}) show how the point group symmetry, which is
present in all local quantities is destroyed for a general
$\ul{k}$-point. Vertex corrections to transport properties vanish
identically on symmetry grounds if we restrict ourselves to the lowest lying
Hund's rule multiplet. Anisotropy effects enter via angular dependent
hybridization matrix elements in the conduction electron self-energy,
Eqn.~(\ref{eqn-sigma-c}). In the case of two doublets  we can
interpret the derived conduction electron self-energy
(\ref{equ-c-self}) as a superposition of contributions of individual
symmetry channels which recovers the proposed semi-phenomenological
extension of the LNCA \cite{Huth95}. Additional maxima in the resistivity
arise naturally from contributions of the different Abricosov-Suhl
resonances from additional CEF levels.

This work has been supported by US Department of Energy, Office of
Basic Energy Science, Division of Material Research and the Deutsche
Forschungsgemeinschaft and in parts by the National Science Foundation
under Grant No.~PHY94-07194. We like to thank the ITP, Santa Barbara for
its hospitality.

\end{document}